\documentclass[preprint2]{aastex}

\usepackage{graphicx}

\shorttitle{Radio galaxy clustering at $z\sim 0.3$}
\shortauthors{M.\ Lacy}

\begin{document}


\title{Radio Galaxy Clustering at $z\sim 0.3$}


\author{Mark Lacy}
\affil{IGPP, L-413, Lawrence Livermore National Laboratory, 
Livermore, CA 94550 and Department of Physics, University 
of California, 1 Shields Avenue, Davis, CA 95616}
\email{mlacy@igpp.ucllnl.org}

\begin{abstract}

Radio galaxies are uniquely useful as probes of large-scale 
structure as their uniform identification with giant elliptical 
galaxies out to high redshift means that the evolution of their bias 
factor can be predicted. As the initial stage in a project to study
large-scale structure with radio galaxies we have performed a small 
redshift survey, selecting 29 radio galaxies in the range $0.19<z<0.45$
from a contiguous 40 deg$^2$ area of sky. We detect significant 
clustering within this sample. The amplitude of the two-point correlation 
function we measure is consistent with no evolution from the local ($z<0.1$)
value. This is as expected in a model in which radio 
galaxy hosts form at high redshift and thereafter obey a continuity 
equation, although the signal:noise of the detection is too low to 
rule out other models. Larger surveys out to $z\sim 1$ 
should reveal the structures of superclusters at intermediate 
redshifts and strongly 
constrain models for the evolution of large-scale structure.
\end{abstract}



\section{Introduction}

Powerful radio sources are almost exclusively associated with giant elliptical 
galaxies, and appear to be in richer than average environments (e.g.\
Hill \& Lilly 1991). 
This suggests that they should be more biased tracers of the mass distribution
than normal galaxies. A study of the clustering of local ($z<0.1$) radio 
galaxies by Peacock \& Nicholson (1991) showed that this was indeed the case, 
with radio galaxies having a cross correlation function of the usual form
(assumed throughout this paper), 
$\xi_{gg}=(r/r_0)^{-1.8}$, with a 
correlation length of $r_0=11 h^{-1}$ Mpc. 
\footnote{We assume a cosmology with 
$\Omega_{\rm M} =1$, $\Omega_{\Lambda} =0$
and $H_0=100 h^{-1} {\rm kms^{-1} Mpc^{-1}}$ throughout}
This can be compared to $r_0=5.7h^{-1}$Mpc for normal 
galaxies (Loveday et al.\ 1992); $r_0=4.5h^{-1}$ for IRAS-selected galaxies
(Fisher et al.\ 1994), and $r_0=14.3h^{-1}$Mpc for rich clusters of galaxies 
(Dalton et al.\ 1994). Radio galaxies thus cluster with a strength
intermediate between normal galaxies and clusters (Bahcall \& Chokshi 1992). 
The associated cosmological bias factor is 
lower than for clusters, but 
about twice as high as that for IRAS-selected galaxies (Peacock \& Dodds 1994).

Studies of the angular correlation function of radio sources by 
Magliocchetti et al.\ (1999) and Cress \& Kamionkowski (1998) 
show that the data from radio surveys such as FIRST are consistent with
little evolution in the clustering amplitude in the range 
$0<z\stackrel{<}{_{\sim}}1$. 
However, these conclusions are based on extrapolating the luminosity 
functions of Dunlop \& Peacock (1990; hereafter DP), to the faint flux 
densities near the limit of the FIRST survey, $S_{1.4}\sim 3$ mJy. At these 
levels the DP luminosity functions are constrained by source count 
data only, as direct redshift surveys were only available at 
$S_{1.4} \stackrel{>}{_{\sim}} 200\,$mJy. In particular, at faint flux levels 
the radio source population is a mix of nearby star-forming galaxies and 
AGN-powered radio sources with a range in redshifts from 0 to $>$4 
(Condon et al.\ 1998), which have quite different clustering properties. 
It is therefore important to test the results of the angular correlation 
function studies with direct measurements of clustering from radio galaxy 
redshift surveys. 

Studies of the clustering of radio-quiet AGN seem to show 
a generally similar correlation length, but there is a wide range in 
estimates of $r_0$ from different samples. This can probably be explained 
if the correlation function depends both on redshift and AGN luminosity
[e.g.\ Sabbey et al.\ (2000), La Franca, Andreani \& Christiani
(1998)]. 


Magliocchetti et al.\ (1999) discuss theoretical predictions for the 
evolution of the two-point correlation function of radio sources. 
Perhaps the most appropriate case to take is that 
where radio galaxies form at high redshift ($z\gg 1$).
We can trace the evolution of the 
host population out to $z\sim 3$, and find that the hosts vary little 
with redshift, apart from some passive evolution. The host magnitudes
are also only weakly dependent on radio luminosity 
(Lacy, Bunker \& Ridgway 2000).
Hence uncertainties in the evolution in the bias factor are
unlikely to be as important an issue for radio galaxies as they are 
for normal galaxies or radio-quiet quasars. 
Fry (1996) shows that in this ``galaxy conservation'' scenario, the 
bias factor increases with redshift according to $b(z)=1+(b_0-1)(1+z)$, 
where $b_0$ is the bias factor at the present epoch. This is because 
fluctuations in the galaxy density field are fixed at the epoch of formation, 
but the fluctuations in 
the matter density field grow with time. The decrease in the bias factor with 
time is mostly compensated for by the clustering of matter under gravity,
for which the growth factor $D(z)=(1+z)^{-1}$ for an $\Omega_{M}=1$, 
$\Omega_{\Lambda}=0$ cosmology.
The two-point correlation function, $\propto D^2(z)b^2(z)$, 
should therefore show little evolution.

To test this model, and to examine the nature of intermediate redshift 
superclusters, 
we therefore decided to begin a survey of large-scale structure at
moderate redshifts ($z\sim 0.2-0.65$). This paper describes the initial result 
from this survey and also the prospects for future surveys.

\section{Survey strategy and observations}

We tried to optimize the survey to detect 
supercluster-scale objects at $z\sim 0.4$. We therefore picked 
a point on the radio luminosity function where the space density of objects 
close to the flux limit would be $\sim 10^{-5}h^3$Mpc$^{-3}$, thus obtaining 
several objects in structures of linear sizes $\sim 100h^{-1}$ Mpc. 
This corresponds to a radio luminosity of $\approx 2\times 10^{23}h^{-2}$
WHz$^{-1}$sr$^{-1}$, or a flux limit of $\approx 20$mJy at 1.4 GHz. This 
is comfortably above the completeness limits of the FIRST and NVSS
radio surveys (Becker et al.\ 1995; Condon et al.\ 1998).

Initial selection was made from the NVSS catalogue, each NVSS source was 
examined in FIRST to check for confusion, and to estimate the position 
of the identification.
This technique combines the sensitivity to extended flux of the 
NVSS survey with the positional accuracy of FIRST. The sample 
discussed in this paper consists of 322 objects within 
right ascension and declination ranges of 
$01^{h} < {\rm R.A.} < 01^{h} 48 ^{m}$ and 
$-02^{\circ} < {\rm Dec.} < 01^{\circ} 20^{'}$ (an area of 
$\approx 40 \, {\rm deg^2}$).

Identifications were made on the UK Schmidt Telescope plates using the
Cambridge APM. The plates were approximately calibrated onto $R$-band
using CCD images of star fields close to each of the plate centres. 
Initially, only the 34 objects classed as non-stellar in $R$ and 
with $17.0\leq R \leq 20.2$ were considered for spectroscopy
(one of these turned out to be a misclassified high redshift quasar). 
Six stellar objects were also selected later as checks on the APM classifier.
For compact sources, or those with identifiable central components the 
probability of a chance misidentification is low. In the survey region
the density of objects with $17.0\leq R \leq 20.2$ is $\approx 0.6$ 
arcmin$^{-2}$. The error on a FIRST position is $\stackrel{<}{_{\sim}}2^{''}$,
so there is only an $\approx 0.2$\% chance of a misidentification. For
double sources with no central component, (18 out of
36 objects) the probability of a misidentification is larger. For these objects
we followed the prescription of Lacy et al.\ (1993) by searching an ellipse
with a major axis equal to the radio source size $d$ and minor axis of $d/2$. 
The number of objects falling into the search region by chance, $N$ 
is given in Table 1, and is $<1$ for all our objects, and 
$\ll 1$ for most of them.
There are very unlikely to be any $z<0.7$ quasars in the sample
-- the fraction of quasars in complete samples drops rapidly at radio 
luminosities below $L_{1.4}\sim 10^{25}h^{-2}$WHz$^{-1}$sr$^{-1}$ 
(Willott et al.\ 2000).

Spectra were obtained on the Shane
3-m Telescope at Lick Observatory on 1999 October 13-14, 
1999 November 12-14 and 1999 December 11-12 and at the Nordic Optical Telescope
(NOT) on 2000 January 5 (all dates UT). The plate calibration 
observations were made on the Shane on 1999 September 15. The Kast 
spectrograph was used for all observations on the Shane, and the Andalucia 
Faint Object Spectrograph (ALFOSC) for the NOT observations. 
Redshifts were determined from emission
lines where present, otherwise absorption features were matched both by 
eye and by a cross-correlation of the spectrum with 
that of a nearby elliptical; the dominant error sources were the accuracy 
of the wavelength calibration, and centroiding of spectral features, both
$\approx 0.3$nm, resulting in a typical 
redshift error of $\pm 0.001$. Full details and the spectra will be presented
in a future paper.

\begin{deluxetable}{lrrcl}
\footnotesize
\tablecaption{Radio galaxies in the sample \label{tbl-1}}
\tablewidth{0pt}
\tablehead{
\colhead{Galaxy}&\colhead{$z$}&\colhead{$R$}&\colhead{Cl}&\colhead{$N$}}
\startdata
010151+011414&0.438&19.0&nn&0.015\\
010236$-$005007&0.243\tablenotemark{a}&17.5&nn&0.002\\
010242$-$005032&0.244\tablenotemark{a}&17.4&nn&0.002\\
010403$-$002437&0.280\tablenotemark{a}&17.2&nn&0.002\\
010456+000422&0.276&18.4&nn&0.13\\
010454+000400&0.280&17.6&nn&0.005\\
010610+005153&0.262&18.2&nn&0.002\\
010620+000843&0.271&18.4&nn&0.027\\
011012$-$004747&0.565&19.6&sn&0.002\\
011341+010609&0.281&18.0&nn&0.002\\
011425+002932&0.355&18.5&nn&0.002\\
011429+000037&0.389&18.8&nn&0.017\\
011527$-$000000&0.381&20.1&nn&0.013\\
011811+010211&0.278&17.3&nn&0.002\\
012012$-$003837&0.236&17.2&nn&0.002\\
012020$-$002124&0.354&19.0&nn&0.002\\
012030$-$001950&0.352&18.7&nn&0.002\\
012101+005100&0.237&18.1&nn&0.033\\
012156$-$002930&0.437&19.2&nn&0.23\\
012411+004050&0.366&18.9&ns&0.002\\
012927+000524&0.392&18.8&nn&0.002\\
013352+011345&0.308&17.8&nn&0.002\\
013506+011913&0.358&18.2&nn&0.004\\
013602+001016&0.344&17.8&nn&0.007\\
013807$-$01452&0.640&18.8&ss&0.010\\
013942$-$000618&0.197\tablenotemark{a}&17.9&ns&0.002\\
014057+001053&0.523&19.9&nn&0.002\\
014213$-$001326&0.526&19.2&nn&0.002\\
014227+001139&0.326&17.7&nn&0.015\\
014251$-$000028&0.272&18.2&nn&0.30\\
014300$-$000245&0.428&19.0&sn&0.43\\
014317$-$011859&0.520&19.5&nn&0.013\\
014347+004546&0.218&18.0&nn&0.002\\
014714+005834&0.638&19.0&nn&0.020\\
014714$-$012603&0.158&17.0&nn&0.55\\
014752+000658&0.448&19.3&nn&0.002\\
\enddata


\tablenotetext{a}{Redshift from other FIRST follow-up projects}

\tablecomments{
Entries in the column labelled `Cl' are `nn' for non-stellar on both 
UKST plates, `ns' for non-stellar on the
$R$ plate and stellar on the $B$ plate, `sn' for stellar on the  $R$
plate and non-stellar on the $B$ plate, and `ss' for stellar on both. $N$
is the expected number of objects with $17\leq R \leq 20.2$
falling into the search region by chance.}
\end{deluxetable}

\section{Analysis}

To measure the two-point correlation function simulated surveys were 
constructed with the same selection function as the original 
survey, but with randomly assigned coordinates. The redshift selection function
was estimated by using the scatter in the observed $R-z$ relation to estimate
the probability of an object with redshift $z$ having a magnitude within
the selected range (Fig.\ 1). A least deviation fit to the $R-z$ relation 
for $z<0.7$ radio galaxies from the 8C-NEC redshift survey 
(Lacy et al.\ 1999) was used to establish a mean $R-z$ relation:
\[ R = 21.3 - 5.9 \, {\rm lg}z \;. \]
This is close to that derived for the 
100-times radio-brighter 3C sample (Eales 1985). 
The scatter about this relation was measured to be 0.57 mag,
using the actual $R$ and $z$ values obtained for sources in the survey. 
A selection function was then constructed 
by taking the luminosity-density evolution model of DP
and integrating to obtain a redshift distribution. This was then multiplied
by the effect of the photometric selection, which was modelled as a pair 
of oppositely-tailed error functions with a width in log redshift 
corresponding to the scatter in the $R-z$ relation, and half-power points 
corresponding to the predicted redshifts for objects at the bright and 
faint magnitude limits of the survey. 

\begin{figure}[ht]
\centering
\includegraphics[scale=0.3]{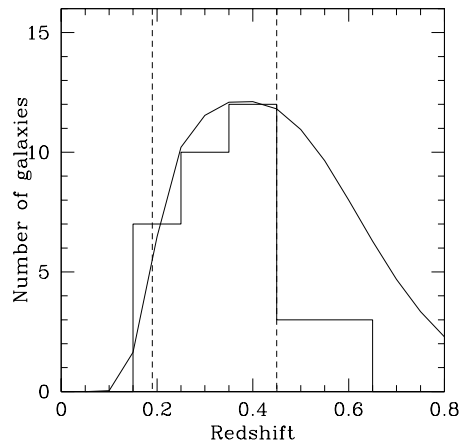}
\caption{The selection function for the sample. The observed
redshift distribution is indicated by the histogram, the theoretical 
selection function by the solid line. The dashed lines indicate the 
redshift range over which we believe our completeness is good
($0.19<z<0.45$).\label{selfn}}
\end{figure}

In practice an upper cutoff of $z=0.45$ was placed on the sample.
Up to this redshift the agreement with 
the predicted redshift distribution is good, with 28.7 objects predicted 
in the range $0.19<z<0.45$ compared to 29 observed.
Above $z\approx 0.45$, however, the predicted
selection function and the observed redshift distribution diverge rapidly
(Fig.\ \ref{selfn}), and of 47.5 objects predicted to be present in the 
redshift range $0.19<z<0.65$, only 36 are found. This could be due either to 
the mistaking of high redshift galaxies for quasars by the 
APM classification program, a problem with the assumed luminosity function 
or a genuine underdense region. 

We next estimate the possible effect of incompleteness due to 
misclassification at $z<0.45$.
There are 12 objects within the survey selection criteria but with stellar 
identifications on the UKST plates for which spectra have not yet been 
obtained. With one exception all have radio morphologies more consistent 
with being quasars than galaxies (i.e.\ unresolved, triple or core-halo
structures). Of the 13 $R$ plate stellar identifications in the range
$17.0\leq R\leq 20.2$ for which we have 
spectra in this region of sky, from both this work and the FIRST spectroscopic 
database, three are in fact galaxies, but only one is at 
$z<0.45$. We therefore think it very unlikely that more than 2-3
objects below $z\sim 0.45$ are missing from the
sample other than those accounted for in the selection function.

\begin{figure}[ht]
\epsscale{1.0}
\plotone{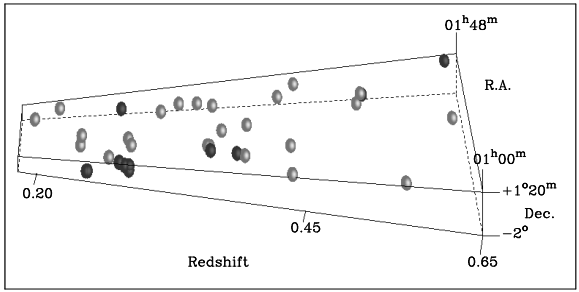}
\figcaption[lacy_fig2.ps]{A representation of the three dimensional 
distribution of galaxies in the sample. FRI radio sources are shown 
in dark grey, FRII and compact sources (i.e.\ those barely 
resolved or unresolved by FIRST) 
are shown as light grey (classifications were made on the basis of radio 
morphology, and only for sources well-resolved by FIRST). 
The diameter of the spheres is 20$h^{-1}$
comoving Mpc. \label{3dfig}}
\end{figure}

In Fig.\ \ref{3dfig}
we show a three dimensional representation of the 
survey volume. Although this figure needs to be interpreted with caution as
the space density of objects is dropping with redshift, there is a 
clump at $z\approx 0.28$ and R.A. $\sim 01^{h}05^{m}$, which seems to be 
comprised mostly of FRI sources. There is also a looser association at 
$z\approx 0.35$, 
and a group of three at $z\approx 0.52$. The sizes of the associations
seem to be $\sim 30-100 \, h^{-1}{\rm Mpc}$, comparable to low redshift 
superclusters.

We have estimated the statistical significance of our detection of clustering 
in the complete $0.19<z<0.45$ sample (median redshift $\approx 0.3$) 
by binning the distribution
of pair separations in 10$h^{-1}$ Mpc bins, and comparing with the 
mean distribution 
obtained from 10000 Monte-Carlo simulations of the survey using a $\chi^2$
test. This gives the probability of our distribution arising by chance
as $4 \times 10^{-6}$ ($\chi^2=112$ with 53 degrees of freedom). 
However, a large part of the $\chi^2$ is contributed 
by the first bin which contains two pairs with separations $<5 h^{-1}$Mpc. 
This is close enough that the members of each pair could be 
in the same galaxy cluster.
We therefore removed these two from the first bin and recalculated
the $\chi^2$ statistic, this gave a probability of 0.006 ($\chi^2=82$ with 
53 degrees of freedom).

To estimate the two-point correlation function, 
the numbers of data-data (DD), data-random (DR) and 
random-random (RR) pairs were measured from the data and the simulations
and the two-point correlation function calculated according to the formula 
espoused by Landy \& Szalay (1993):

\[ \xi(r) = \frac{DD - 2DR + RR}{RR}. \]

\begin{figure}[ht]
\epsscale{0.8}
\plotone{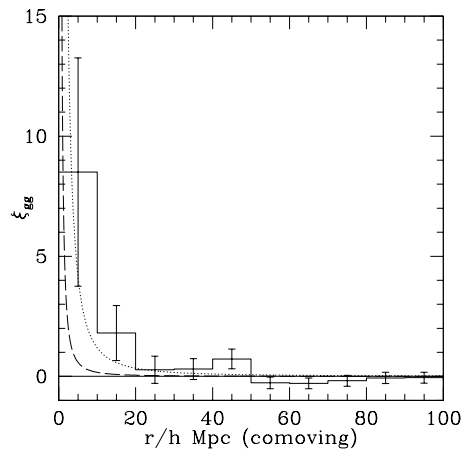}
\figcaption[lacy_fig3.ps]{The two-point correlation function of the 
$0.19\leq z\leq 0.45$ radio galaxies.
Error bars are Poisson. The dotted line is the local radio galaxy 
correlation function from Peacock \& Nicholson (1991), the dashed line
the normal galaxy correlation function at $z\sim 0.3$ (Small et al.\ 1999). 
\label{bgg}}
\end{figure}

The results for the complete $0.19<z<0.45$ sample are shown in 
Fig.\ \ref{bgg}. The correlation function was fit over the range 
0-100 $h^{-1}$Mpc, the best value for $r_0$ was $\approx 17h^{-1}$Mpc, 
with a range of 5--24$h^{-1}$Mpc over which 
the probability of obtaining the $\chi^2$ was $>$5 \%  
(assuming Poisson errors). 

As a check on the effect of possible incompleteness
we added three galaxies (the largest number we expect based on the discussion
above) to the sample with redshifts drawn at random from the 
selection function, and with random sky positions within the survey region.
This reduced $r_0$ to 14$h^{-1}$Mpc. Incompleteness is thus unlikely 
to affect our estimate by more than the random error.

\section{Conclusions and future surveys}

We have succeeded in  
developing an effective method for studying the clustering of 
moderate redshift radio galaxies directly, and 
have detected clustering of radio galaxies at $z\approx 0.3$. 
The amplitude of the cross-correlation 
function we measure is consistent with that for radio
galaxies locally. This is as expected in the simple model discussed in the 
introduction, in which radio source hosts evolve little with redshift, and is 
higher than that for normal galaxies at $z\sim 0.3$, for which Small et 
al.\ (1999) measure $r_0=3.7\, h^{-1} {\rm Mpc}$. 
At present, however, the small size of 
our survey prevents us ruling out all but the most extreme evolution in 
the correlation function. Expansion of this survey to 
$\stackrel{>}{_{\sim}} 100$ redshifts will allow a measurement of the 
two-point correlation function to be made which has comparable accuracy to 
that for normal galaxies at these redshifts. 
The relatively large volume probed by a larger survey will allow us to define 
a sample of superclusters and to examine their structures and the 
evolution of those structures to the present day. 

The discovery of redshift 
clustering in deep pencil beam galaxy 
surveys, e.g. that of the HDF (Cohen et al.\ 2000) has raised the 
possibility that large-scale structures continue to be present in the 
Universe at least to $z\sim 1$. 
Therefore the evolution of these 
structures should place interesting constraints on cosmology.
Crucially, however, because we can trace the 
evolution of radio galaxy hosts to $z\sim 3$, we can, in principle,
predict how the
bias should evolve with redshift, removing an important uncertainty from 
the interpretation of the results of correlation function studies. 

\acknowledgments

We thank the staff at the Lick Observatory and NOT for their help with the 
observations, in particular the night assistants on the Shane, Andy, Wayne 
and Keith,  
and our support astronomers Elinor Gates and Hugo Schwartz. We also thank
Dan Stern, Susan Ridgway, Sally Laurent-Meulheisen and Margrethe Wold for
help with various aspects of the project. We thank Bob Becker, Mike 
Brotherton, Michael Gregg, and Sally Laurent-Meuleisen for spectra
obtained from the FIRST spectroscopic database. We also thank Mike 
Irwin for the APM service, and Devinder Sivia for 
the {\sc pgxtal} software. 
This work was performed under the auspices of the U.S. Department of
Energy by University of California Lawrence Livermore National
Laboratory under contract No. W-7405-Eng-48, with support from 
NSF grants AST-98-02791 and AST-98-02732. 
The NOT is operated on the Island of La Palma jointly by Denmark, Finland, 
Iceland, Norway and Sweden, in the Spanish Observatorio del Roque de los 
Muchachos of the Instituto de Astrofisica de Canarias.





\begin{thebibliography}{}
\bibitem[Bahcall \& Chokshi (1992)]{BC92} 
Bahcall N.A., Chokshi A., 1992, ApJ, 385, L33
\bibitem[Becker, White \& Helfand (1995)]{FIRST} Becker R.H., White R.L., 
Helfand D.J., 1995, ApJ, 450, 559
\bibitem[Cohen et al.\ (2000)]{Coh00} Cohen J., Hogg D., Blandford R., Cowie
L., Hu E., Songaila A., Shopbell P., Richberg K., 2000, ApJ in press
(astro-ph/9912048)
\bibitem[Condon et al.(1998)]{NVSS} Condon J.J., Cotton W.D., Greisen E.W., 
Yin Q.F., Perley R.A., Taylor G.B., Broderick J.J., 1998, AJ, 115, 1693
\bibitem[Cress \& Kamionkowski (1998)]{CK98} Cress C.M., Kamionkowski M., 
1998, MNRAS, 297, 486
\bibitem[Dalton et al.\ 1994]{Dal94} Dalton G.B., Croft R.A.C., Efstathiou
G.P., Sutherland W.J., Maddox S.J., Davis M., 1994, MNRAS, 271, L47 
\bibitem[Dunlop \& Peacock (1990)]{DP90} Dunlop J.S., Peacock J.A., 1990, 
MNRAS, 247, 19 (DP)
\bibitem[Eales (1985)]{EaIII} Eales S.A., 1985, MNRAS, 217, 179
\bibitem[Fisher et al.\ (1994)]{Fis94} Fisher K.B., Davis M., Strauss M.A., 
Yahil A., Huchra J., 1994, MNRAS, 266, 50
\bibitem[Fry (1996)]{Fry96} Fry J.N., 1996, ApJ, 461, L65
\bibitem[Hill \& Lilly (1991)]{HL91} Hill G.J., Lilly S.J., 1991, ApJ, 367, 1
\bibitem[Lacy et al.\ (1993)]{PII} Lacy M., Hill G.J., Kaiser M.E., Rawlings
S., 1993, MNRAS, 263, 707
\bibitem[Lacy et al.\ (1999)]{PIII} Lacy M., Kaiser, M.E., Hill G.J., 
Rawlings S., Leyshon G., 1999, MNRAS, 308, 1087
\bibitem[Lacy, Bunker \& Ridgway (2000)]{LBR00} Lacy M., Bunker A.J., 
Ridgway S.E., 2000, AJ, in press (astro-ph/0003290)
\bibitem[La Franca et al.\ (1998)]{LAC98} La Franca F., Andreani P., 
Christiani S., 1998, ApJ, 497, 529
\bibitem[Landy \& Szalay (1993)]{LS93} 
Landy S.D., Szalay A.S., 1993, ApJ, 412, 64
\bibitem[Loveday et al.\ (1992)]{Lov92} Loveday J., Efstathiou G.P., 
Peterson B.A., Maddox S.J., 1992, ApJ, 400, L43 
\bibitem[Magliocchetti et al.\ (1999)]{Mag99} Magliocchetti M., Maddox S.J., 
Lahav O., Wall J.V., 1999, MNRAS, 306, 943
\bibitem[Peacock \& Dodds (1994)]{PD94} Peacock J.A., Dodds S.J., 
1994, MNRAS, 267, 1020
\bibitem[Peacock \& Nicholson (1991)]{PN91} Peacock J.A., 
Nicholson D., 1991, MNRAS, 253, 307
\bibitem[Sabbey et al.\ (2000)]{Sab00} Sabbey C.N., et al., ApJ, submitted
(astro-ph/9912108)
\bibitem[Small et al.\ (1999)]{Sma99} Small T.A., Ma C.-P., Sargent W.L.W., 
Hamilton D., 1999, ApJ, 524, 31
\bibitem[Willott et al.\ (2000)]{QF} Willott C., Rawlings S., Blundell K.M., 
Lacy M., 2000, MNRAS, in press (astro-ph/0003461)
\end{thebibliography}
\end{document}